\documentclass[aip,cha,reprint,amsmath,amssymb,nobibnotes,showkeys,floatfix]{revtex4-2}
\usepackage[T1]{fontenc}

\usepackage{graphicx}
\usepackage{mathrsfs}
\usepackage{amsthm}
\usepackage[hidelinks]{hyperref}
\hypersetup{
  pdftitle={Inferring Private Network Structure from Public Opinions in Two Layer Opinion Dynamics},
  pdfauthor={Xiaoqun Wu, Guanfu Zhong, Yutong Christina Wang, Ruiwu Niu}
}
\newtheorem{theorem}{Theorem}
\newtheorem{proposition}{Proposition}
\newcommand{\modelFigure}{%
  \begin{figure*}[t]
    \centering
    \includegraphics[width=\linewidth]{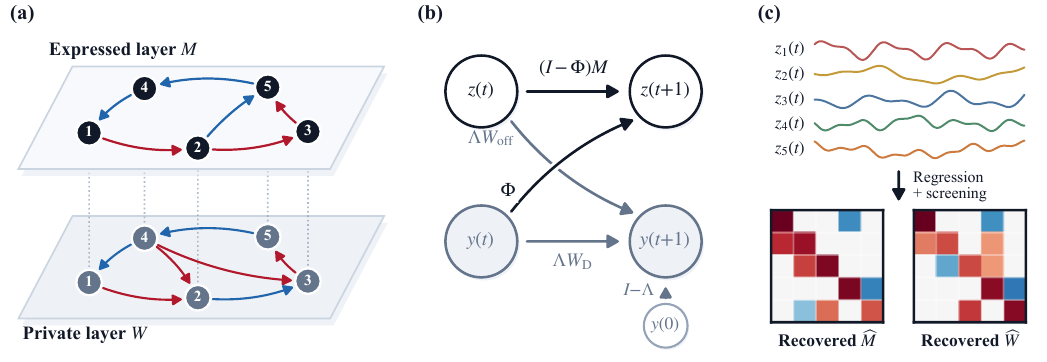}%
    \caption{Signed expressed--private opinion model and reconstruction. (a) Five individuals have an expressed layer $M$ and a private layer $W$. Red and blue arrows denote positive and negative influence, respectively. Dotted lines connect the same individual across layers. (b) The coupled updates link the observed expressed opinions $z(t)$ to the unobserved private opinions $y(t)$ and their initial values $y(0)$. (c) Once the reduced coefficients are identified, trajectory compatibility screens the candidate rows. The matrices are uniquely recovered when one assignment remains at every node.}
    \label{fig:1}
  \end{figure*}%
}

\newcommand{\pairFigure}{%
  \begin{figure*}[t]
    \centering
    \includegraphics[width=\linewidth]{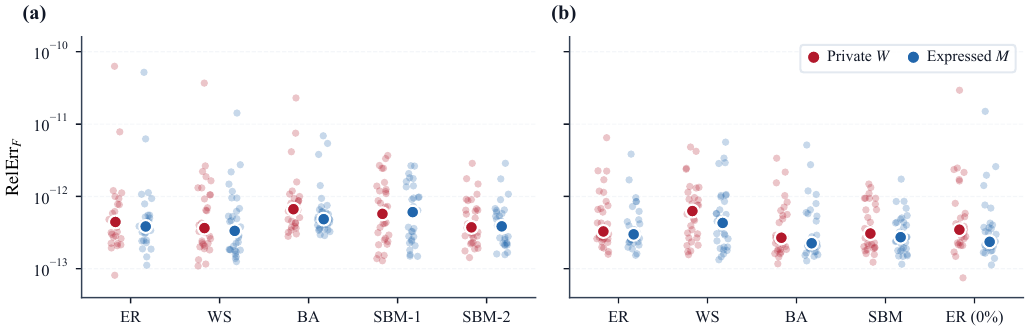}%
    \caption{Relative Frobenius errors for the private layer $W$ (red) and the expressed layer $M$ (blue). (a) Five partially overlapping graph pairs: ER, WS, BA, and two four-block models, SBM-1 and SBM-2. (b) Independent private layers paired with each displayed expressed topology, plus an ER control with disjoint edge sets. Points show the 30 independent trials per configuration and outlined circles mark medians; both layers achieve $\mathrm{F1}=1.000$ and $\mathrm{SignAcc}=1.000$ throughout.}
    \label{fig:2}
  \end{figure*}%
}

\newcommand{\scaleFigure}{%
  \begin{figure*}[t]
    \centering
    \includegraphics[width=\linewidth]{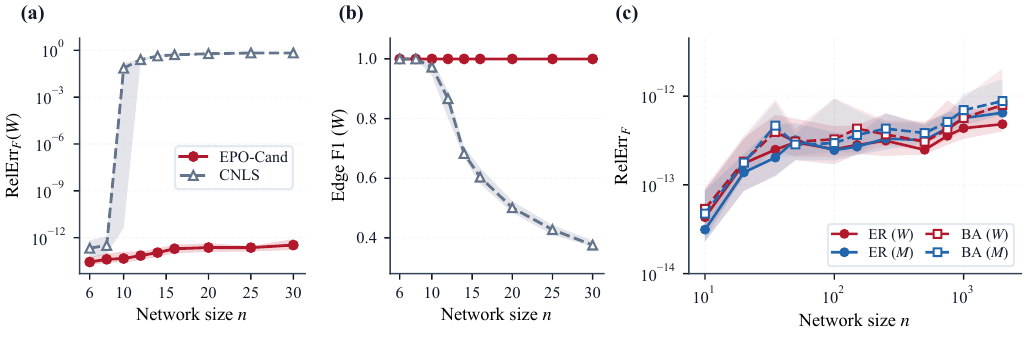}%
    \caption{Comparison with CNLS and multi-scale accuracy versus network size $n$. (a) Private-layer $\mathrm{RelErr}_F(W)$ and (b) edge F1 comparing EPO-Cand against multi-start CNLS on ER networks for $n=6$ to 30. (c) Two-layer $\mathrm{RelErr}_F$ of EPO-Cand on sparse ER and BA networks from $n=10$ to 2,000. Curves show medians across 30 independent trials at each network size and shaded bands indicate 25\%--75\% interquartile ranges. EPO-Cand attains $\mathrm{F1}=1.000$ and $\mathrm{SignAcc}=1.000$ in all 270 trials in (a),(b) and in all 658 of 660 uniquely resolved trials in (c).}
    \label{fig:3}
  \end{figure*}%
}

\newcommand{\noiseFigure}{%
  \begin{figure}[!t]
    \centering
    \includegraphics[width=.92\linewidth]{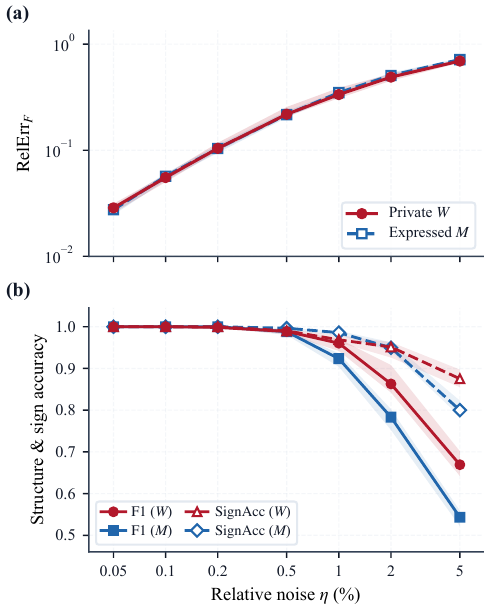}%
    \caption{Reconstruction accuracy under mild Gaussian observation noise on 50-node ER networks with $K=400$ trajectories and $T=3$ updates. (a) Relative Frobenius errors $\mathrm{RelErr}_F$ and (b) edge F1 and SignAcc at threshold $\tau=0.025$ for both layers across seven relative noise levels $\eta$. Curves show medians over 30 independent trials at each noise level and shaded bands indicate 25\%--75\% interquartile ranges.}
    \label{fig:4}
  \end{figure}%
}

\newcommand{\topologyFigure}{%
  \begin{figure}[!t]
    \centering
    \includegraphics[width=.76\linewidth]{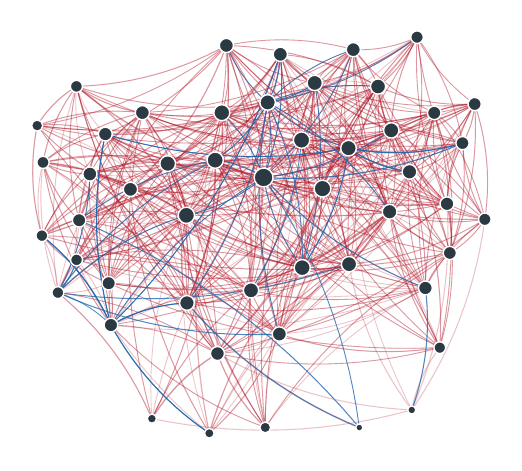}%
    \caption{Signed network topology of the Top-50 active core subgraph extracted from the SNAP Bitcoin OTC trust network. Node area increases monotonically with total directed degree, with higher-degree hubs positioned toward the center. Red and blue curved arcs denote positive trust ratings and negative distrust ratings, respectively, drawn on a unified edge-width scale with opposite curvatures separating reciprocal directed ties.}
    \label{fig:5}
  \end{figure}%
}

\newcommand{\bitcoinFigure}{%
  \begin{figure*}[t]
    \centering
    \includegraphics[width=\linewidth]{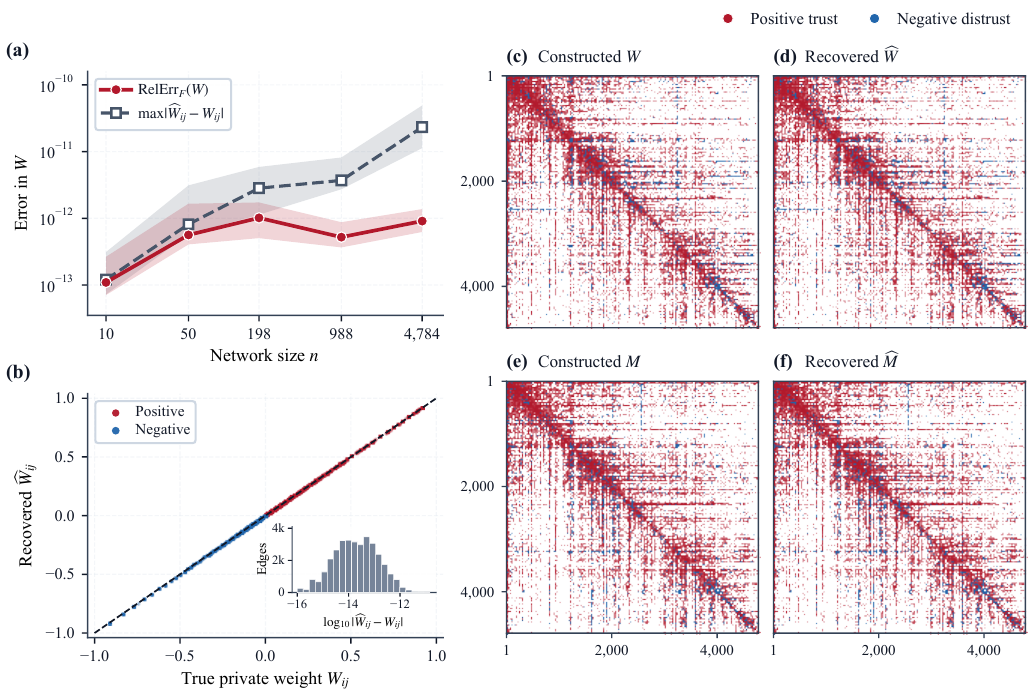}%
    \caption{Reconstruction on the Bitcoin OTC trust topology. (a) Median $\mathrm{RelErr}_F(W)$ and $\max|\widehat W_{ij}-W_{ij}|$ with 25\%--75\% interquartile bands over 30 independent trials at each of the five subgraph scales, with $\mathrm{F1}=1.000$ and $\mathrm{SignAcc}=1.000$ in every trial. (b) True versus recovered private edge weights on the 4,784-node core with the log-error distribution inset. (c),(d) Constructed and recovered off-diagonal patterns of $W$, and (e),(f) those of $M$, with rows and columns indexing receiving and sending nodes, respectively.}
    \label{fig:6}
  \end{figure*}%
}

\newcommand{\boundaryFigure}{%
  \begin{figure}[!t]
    \centering
    \includegraphics[width=.92\linewidth]{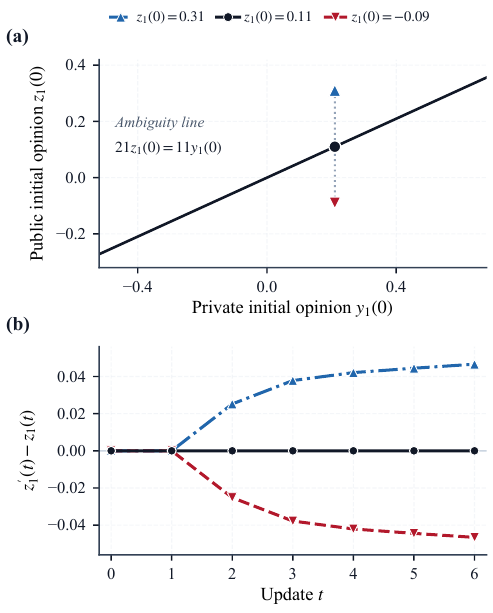}%
    \caption{Two admissible networks with opposite signs for the private edge $2\to1$. (a) The circle lies on the ambiguity line $21z_1(0)=11y_1(0)$; the triangles shift $z_1(0)$ by $\pm0.2$ at fixed $y_1(0)=0.21$. (b) Difference in the two predictions for node 1 after matching the initial public opinions and the first update. All three records identify $H_1,H_0$ from six updates; only the record on the line admits both signs.}
    \label{fig:7}
  \end{figure}%
}

\begin{document}
\title{Inferring Private Network Structure from Public Opinions in Two Layer Opinion Dynamics}
\author{Xiaoqun Wu}
\author{Guanfu Zhong}
\affiliation{Guangdong Provincial Key Laboratory of Visual Media and Multidimensional Intelligence, CSSE, Shenzhen University, Shenzhen, China}
\author{Yutong Christina Wang}
\affiliation{Department of Computer Science and Engineering, Chinese University of Hong Kong, Shatin, Hong Kong, China}
\author{Ruiwu Niu}
\affiliation{Department of Data Science and Digital Innovation, Hong Kong Shue Yan University, Hong Kong, China}
\date{28 November 2026}
\begin{abstract}
Public opinions can reveal how influence is expressed while leaving the network that shapes private opinions ambiguous. We study a signed two-layer model with unknown influence matrices, individual response parameters, and private initial opinions. Eliminating the private states gives a second-order recurrence for the expressed opinions. Its coefficients leave at most two assignments of the effective self-weights at each node. We derive a necessary and sufficient test for an assignment to reproduce a full trajectory. Once the reduced coefficients are identified, one noiseless trajectory selects the unique assignment for almost every joint initial state, provided each individual retains positive anchoring to their initial private opinion. Based on this characterization, we develop a reconstruction procedure that combines linear regression with independent row-wise compatibility tests. Numerical simulations on noiseless expressed opinions across diverse synthetic graph families, independently generated private layers, and networks constructed from empirical signed topologies demonstrate that the proposed method accurately recovers both signed layers and substantially outperforms budgeted multi-start constrained nonlinear least squares as network size grows. Moreover, the procedure scales reliably to sparse networks of up to 2,000 nodes and to a 4,784-node empirical trust topology, achieving near-machine-precision weight recovery. When exact compatibility screening is replaced by minimum-residual selection, the procedure also preserves high edge and sign recovery accuracy under mild observation noise.
\end{abstract}
\maketitle
\begin{quotation}
When only publicly expressed opinions are visible, can they distinguish two networks that assign different influences to the same person? In our model, private memory and the persistence of public expression enter the observed recurrence through two effective self-weights that can exchange places. The first public update and the subsequent trajectory must nevertheless agree on the contribution of the initial private opinion. This requirement tests the two assignments for each individual. Once the public recurrence coefficients are identified and every individual retains some attachment to their initial private opinion, one noiseless trajectory distinguishes all the alternatives for almost every choice of initial private and expressed opinions. A two-person example makes the exception visible. Along one line of initial opinions, two networks produce identical public statements while assigning opposite signs to the same private influence.
\end{quotation}
\section{Introduction}

How a group forms and maintains consensus amid differing opinions among members is fundamental to understanding collective decision-making, social norms, and the evolution of public opinion. When a norm is widely supported in public, group members may systematically overestimate others' approval of it. Prentice and Miller\cite{r1} found in a survey of campus drinking norms that students generally believed they were less comfortable with campus drinking practices than their peers, and their estimates of their friends' attitudes showed a similar bias. In Asch's line-segment judgment experiment, some participants later reported thinking the majority answer was wrong yet still voiced the same answer in front of others\cite{r2}. These two phenomena can reinforce each other in ongoing social interactions: a person's public conformity to the majority can in turn be taken by others as further evidence of genuine agreement. Therefore, to explain the formation of public consensus, we must examine both how people are influenced by others to revise their private judgments and how they decide which positions to express.

Opinion dynamics provides a mathematical framework for studying how interactions among individuals form group-level consensus and disagreement\cite{r3}. DeGroot describes opinion updates as repeated weighted averages: in fixed, strongly connected networks with positive self-weights, individuals starting from different initial positions reach consensus, with each person's initial judgment contributing according to the network's influence weights\cite{r4}. Friedkin and Johnsen further allow individuals to retain attachment to their initial position at each round of interaction, explaining why groups in ongoing communication can still sustain stable disagreements\cite{r5}. Another research line focuses on the sign of interpersonal ties. Altafini uses negative weights to represent antagonistic influences, showing that structurally balanced networks\cite{r6} can reach bipartite consensus under certain conditions, where two camps each achieve internal agreement while holding opposite positions\cite{r7}. Xia et al.\cite{r8} extended this analysis beyond strongly connected graphs, showing that when a strongly connected subnetwork influences the remaining nodes, whether its internal ties are structurally balanced determines the asymptotic opinion distribution across the entire group. These models all represent each individual's opinion by a single state, making it difficult to distinguish a shift in private judgment from an adjustment in public expression.

Modeling private opinions and visible behaviors separately provides a way to analyze this distinction. Xiong and Liu\cite{r9} analyzed discussions across multiple Twitter topics and observed that a single viewpoint can dominate public discussion even before the group reaches uniform consensus; on that basis, they proposed a model separating internal opinions from explicit actions, where individual activity governs whether one speaks out and neighbors are influenced by visible actions. The expressed and private opinion (EPO) model of Ye et al.\cite{r10} assigns two coupled evolving states to each individual and establishes conditions under which the two opinions remain distinct after convergence. In this model, private updates combine an individual's current private opinion and neighbors' expressed opinions while retaining attachment to the initial private belief, whereas public expression balances private judgment against the local public opinions voiced by neighbors. Consequently, the same neighbor's statement can act through two distinct pathways: reshaping an individual's private judgment on the issue or altering what that individual chooses to express in public.

When trust and antagonism coexist in a network, the distinction between these two pathways becomes especially important. Xie et al.\cite{r11} studied convergence and bipartite consensus of EPO dynamics in signed directed networks under relaxed connectivity conditions, analyzing how individual self-confidence affects the gap between final expressed opinions and local group views. He et al.\cite{r12} analyzed the steady-state properties of both opinions in the presence of stubborn individuals and derived a lower bound on the final opinion divergence. Wang et al.\cite{r13} extended the model to sequences of topics, using the final private opinion of one topic as the initial private belief of the next, and gave sufficient conditions for structurally unbalanced networks to reach bipartite consensus when both layers share the same signed topology. This shared-support assumption permits different magnitudes across layers but requires both layers to share identical edge locations and signs. As illustrated in their example, however, a person may openly agree with someone they do not privately trust. Once this shared-support assumption is relaxed, even knowing that a neighbor positively affects an individual's public expression does not determine whether that neighbor influences the individual's private judgment or with what sign; those private relations must instead be identified from observations.

Identifying influence relations from observations is the network reconstruction counterpart to opinion evolution\cite{r14,r15,r16}. Ravazzi et al.\cite{r17} studied the Friedkin--Johnsen model with stochastic interactions, estimating time-lagged opinion correlations from partial node observations across time to reconstruct the edges and weights of sparse influence networks. Which interactions are identifiable from data also depends on the underlying dynamical process. Wang et al.\cite{r18} used binary node state sequences generated by binary contagion or Ising dynamics to reconstruct pairwise and three-body interactions in two-dimensional simplicial complexes. Related dynamical reconstruction frameworks further address stepwise inference of higher-order networks\cite{r19}, deep-learning reconstruction from incomplete observations\cite{r20}, graph-theoretic topology identification of stochastic multilayer networks\cite{r21}, and joint topology inference in the presence of hidden nodes\cite{r22}. In multilayer networks, where cross-layer coupling shapes collective dynamics\cite{r23}, whether observations can separate layers is equally important. Mei et al.\cite{r24} reconstructed the sparse topology of a target layer using states from both layers, where the auxiliary layer need only be measured at a few time instants; related studies likewise leverage dynamics from both layers to identify response-layer topology\cite{r25} or jointly reconstruct connections and separate unknown layers\cite{r26}. Ma et al.\cite{r27}, starting from single-type aggregated node observations, jointly infer the structure and dynamical parameters of each layer under the assumption that all nodes share homogeneous dynamical parameters within each layer. Complementary studies reconstruct topologies from time series affected by hidden variables\cite{r28}, non-Markovian hidden-layer decompositions\cite{r29}, or partial layer observations\cite{r30}. Public opinion records impose a different observation regime: each measurement tracks a known individual's public expression, yet even when all individuals are monitored continuously, the private layer remains entirely unobserved, while individuals can differ in both their susceptibility to social influence and the weight they place on private judgment in public expression.

In empirical research, He et al.\cite{r31} aggregated Weibo comments into community sentiment time series, estimated directed inter-community influence networks, and evaluated predictive accuracy on held-out observations; they also fitted EPO models to trajectories derived from code embeddings to estimate latent trust networks and private opinions\cite{r32}. To interpret an estimated edge as an influence on private judgment, however, one must ask whether this interpretation is unique, which raises a structural identifiability question\cite{r33}. For linear time-invariant networks, Gon\c{c}alves and Warnick\cite{r34} showed that even when input--output data uniquely determine the transfer function, reconstructing dynamical structure among measured states generally requires additional structural information. Weerts et al.\cite{r35}, starting from a parameterized network model class, showed that both the placement of external excitations and the correlation structure of disturbances govern network identifiability. Subsequent control-theoretic studies establish topology reconstruction via constrained Lyapunov equations\cite{r36}, analyze output discernibility of topological variations in linear and switched dynamical networks\cite{r37,r38}, and characterize how excitation allocation, partial measurement, and diffusive coupling determine module and network identifiability\cite{r39,r40,r41,r42}. For signed EPO models, if an alternative private network paired with different expressed weights and private initial opinions reproduces the exact same public record at every step, then even a zero-residual fit cannot distinguish between the two explanations. The identifiability question therefore reduces to an edge-level test: among all admissible models generating the same public trajectories, does a given private edge always exist, and does it always carry the same sign?

We study a signed expressed--private opinion model in which private and expressed influence weights are normalized by row. The layers can differ in their edges, signs, and weights. Given noiseless expressed opinions, which private influences do these observations uniquely determine when both networks, the individual parameters, and the private initial opinions are unknown? Eliminating the private states gives an exact second-order recurrence for the expressed opinions. Once its coefficients are identified, at most two parameter assignments remain at each node. A candidate assignment must account for both the first public update and the subsequent evolution using the same private initial opinion. We derive a necessary and sufficient condition for this agreement. Comparing the assignments that survive determines whether the observations uniquely identify the presence, sign, or weight of each private edge. Figure~\ref{fig:7} gives a two-node case in which distinct signs for the same private influence produce identical expressed opinions for all times.

\section{Problem Statement}\label{sec:problem}
Consider a group of $n$ individuals represented in two network layers, with a known correspondence between the nodes in the two layers. At each discrete time $t$, let $y_i(t)$ and $z_i(t)$ denote the private and expressed opinions of individual $i$, respectively. The corresponding opinion vectors are $y(t),z(t)\in\mathbb{R}^n$. Let $W$ and $M$ denote the influence matrices for the private and expressed layers, respectively. For $i\ne j$, the entries $W_{ij}$ and $M_{ij}$ specify how the expressed opinion of individual $j$ influences the private and expressed opinions of individual $i$, respectively. Both entries refer to the directed edge $j\to i$. In each layer, a zero entry indicates that the edge is absent, and the sign of a nonzero entry indicates whether the influence is positive or negative. The two layers may differ in which edges are present and in the signs and weights of those edges.

Let $W_{\mathrm{D}}=\operatorname{diag}(W_{11},\ldots,W_{nn})$ and $W_{\mathrm{off}}=W-W_{\mathrm{D}}$ denote the diagonal and off-diagonal parts of $W$, respectively. Define $\Lambda=\operatorname{diag}(\lambda_1,\ldots,\lambda_n)$ and $\Phi=\operatorname{diag}(\phi_1,\ldots,\phi_n)$. The private and expressed opinions evolve according to
\begin{align}
y(t+1)&=\Lambda W_{\mathrm{D}}y(t)+\Lambda W_{\mathrm{off}}z(t)+(I-\Lambda)y(0),\label{eq:t1}\\
z(t+1)&=\Phi y(t)+(I-\Phi)Mz(t),\label{eq:t2}
\end{align}
where $I$ is the $n\times n$ identity matrix. In the private update, individual $i$ combines their current private opinion with their neighbors' expressed opinions according to row $i$ of $W$. This sum receives weight $\lambda_i$, and the individual's initial private opinion $y_i(0)$ receives weight $1-\lambda_i$. In the expressed update, individual $i$ assigns weight $\phi_i$ to their current private opinion and weight $1-\phi_i$ to the weighted sum of expressed opinions $(Mz(t))_i$. Both updates use the states at time $t$, as illustrated in Fig.~1.

\modelFigure

For each individual $i$, assume that $0<\lambda_i\leq 1$, $0<\phi_i<1$, and $W_{ii},M_{ii}\geq 0$. The weights in each layer are normalized by row so that $\sum_{j=1}^n|W_{ij}|=1$, $\sum_{j=1}^n|M_{ij}|=1$. Suppose we observe $K$ noise-free trajectories of expressed opinions, with $T_k$ updates in trajectory $k$. The observed data are
\begin{equation}
D=\left\{\bigl(z^{(k)}(0),z^{(k)}(1),\ldots,z^{(k)}(T_k)\bigr)\right\}_{k=1}^{K}.
\label{eq:t3}
\end{equation}
Each vector $z^{(k)}(t)$ contains the expressed opinions of all $n$ individuals at time $t$. The trajectories share the unknown parameters $\Theta=(W,M,\Lambda,\Phi)$ but may have different initial states. The initial expressed opinion vector $z^{(k)}(0)$ is observed as part of trajectory $k$. The initial private opinion vector $y^{(k)}(0)$ and all subsequent private opinions are unobserved.

A parameter set $\Theta$ satisfying the above assumptions is compatible with $D$ if the following condition holds for every trajectory $k$. There must exist an initial private opinion vector $y^{(k)}(0)$ that, together with the observed $z^{(k)}(0)$, yields the entire observed trajectory under $\Theta$. We seek to identify $W$ and $M$ jointly by determining which edge signs and weights are common to all parameter sets compatible with $D$.

\section{Identification and joint reconstruction of public and private influence networks}\label{sec:theory}
\subsection{Reduction to Expressed Opinions}
Let $P,Q,R\in\mathbb{R}^{n\times n}$ denote the effective private self-influence, private interaction, and expressed interaction matrices, respectively:
\begin{equation}
P=\Lambda W_{\mathrm{D}},\quad Q=\Lambda W_{\mathrm{off}},\quad R=(I-\Phi)M.
\label{eq:t4}
\end{equation}
The private initial opinion contributes the constant $c=(I-\Lambda)y(0)\in\mathbb{R}^n$. Equations~\eqref{eq:t1} and~\eqref{eq:t2} become
\begin{equation}
\begin{aligned}
y(t+1)&=Py(t)+Qz(t)+c,\\
z(t+1)&=\Phi y(t)+Rz(t).
\end{aligned}\label{eq:t5}
\end{equation}
For each trajectory generated by~\eqref{eq:t1} and~\eqref{eq:t2},
\begin{equation}
z(t+2)=H_1z(t+1)+H_0z(t)+\Phi c,
\label{eq:t6}
\end{equation}
where the reduced coefficient matrices $H_1,H_0\in\mathbb{R}^{n\times n}$ are
\begin{equation}
H_1=P+R,\qquad H_0=\Phi Q-PR.
\label{eq:t7}
\end{equation}
Define the forward difference $\Delta z(t)=z(t+1)-z(t)\in\mathbb{R}^n$. Then
\begin{equation}
\Delta z(t+2)=H_1\Delta z(t+1)+H_0\Delta z(t).
\label{eq:t8}
\end{equation}
Substituting the private update into $z(t+2)=\Phi y(t+1)+Rz(t+1)$ and using that the diagonal matrices $P$ and $\Phi$ commute gives
\begin{align*}
z(t+2)&=\Phi Py(t)+\Phi Qz(t)+\Phi c+Rz(t+1)\\
&=P\{z(t+1)-Rz(t)\}+\Phi Qz(t)\\
&\quad+\Phi c+Rz(t+1),
\end{align*}
which is~\eqref{eq:t6}. First differencing removes the constant $\Phi c$.

Let $N=\sum_{k=1}^{K}\max(0,T_k-2)$ count the available regression rows. The matrix $X\in\mathbb{R}^{N\times 2n}$ stacks $[\Delta z^{(k)}(t+1)^\top,\Delta z^{(k)}(t)^\top]$ for $0\leq t\leq T_k-3$ in every trajectory. Whenever $\operatorname{rank}(X)=2n$, ordinary least squares uniquely recovers $(H_1,H_0)$ in exact arithmetic.

Write $\operatorname{sign}:\mathbb{R}\to\{-1,0,1\}$ for the sign function. Because $P$ is diagonal and $1-\phi_i>0$, the expressed network has the immediate off-diagonal relation
\begin{equation}
\begin{aligned}
(H_1)_{ij}&=(1-\phi_i)M_{ij},\\
\operatorname{sign}(M_{ij})&=\operatorname{sign}((H_1)_{ij}),\qquad i\ne j.
\end{aligned}\label{eq:t9}
\end{equation}
Thus the reduced coefficients determine the signed support outside the diagonal of $M$. Recovering $W$ and the node parameters requires resolving the candidate decomposition.

\pairFigure

\subsection{Candidate Decomposition}
Let $p_i=P_{ii}\in[0,\lambda_i]$ and $r_i=R_{ii}\in[0,1-\phi_i]$ denote the two effective self weights at node $i$. Since $Q_{ii}=0$, Eq.~\eqref{eq:t7} gives
\begin{equation}
(H_1)_{ii}=p_i+r_i,\qquad (H_0)_{ii}=-p_ir_i.
\label{eq:t10}
\end{equation}
Hence $p_i$ and $r_i$ are the two unordered roots of
\begin{equation}
x^2-(H_1)_{ii}x-(H_0)_{ii}=0,\qquad x\in\mathbb{R}.
\label{eq:t11}
\end{equation}
For each assignment of the two roots to $p_i$ and $r_i$, the reduced coefficients and the row normalization of $W$ and $M$ determine the remaining parameters for node $i$:
\begin{align}
\phi_i&=1-r_i-\sum_{j\ne i}|(H_1)_{ij}|,\label{eq:t12}\\
R_{ii}&=r_i,\quad R_{ij}=(H_1)_{ij}\quad(i\ne j),\label{eq:t13}\\
M_{ij}&=\frac{R_{ij}}{1-\phi_i},\qquad j=1,\ldots,n,\label{eq:t14}\\
Q&=\Phi^{-1}(H_0+PR),\label{eq:t15}\\
\lambda_i&=p_i+\sum_{j\ne i}|Q_{ij}|,\quad W_{ii}=\frac{p_i}{\lambda_i},\quad
 W_{ij}=\frac{Q_{ij}}{\lambda_i}\quad(j\ne i).\label{eq:t16}
\end{align}

\begin{theorem}\label{thm:reduced}
Suppose that $H_1$ and $H_0$ are known exactly and that the model satisfies the parameter bounds and row normalization conditions stated in Sec.~\ref{sec:problem}. At each node $i$, assigning the two roots to $p_i$ and $r_i$ in either order yields at most two distinct admissible decompositions. If the roots coincide, only one assignment is possible. The admissible parameter tuples that realize $(H_1,H_0)$ are precisely those obtained by choosing an admissible assignment independently for each row and applying the reconstruction formulas above.
\end{theorem}

\begin{proof}
Every admissible model realizing $(H_1,H_0)$ must satisfy~\eqref{eq:t10}, so its $p_i,r_i$ are an assignment of the two roots in~\eqref{eq:t11}. Row normalization then fixes $\phi_i,M_{i:},Q_{i:},\lambda_i,W_{i:}$ by~\eqref{eq:t12}--\eqref{eq:t16}, where the subscript $i:$ denotes row $i$. This proves that no admissible decomposition is omitted. Conversely, substitution of any admissible constructed row into~\eqref{eq:t7} recovers the given reduced coefficients. An assignment is rejected if it violates the parameter intervals or nonnegative diagonals; the construction enforces row normalization. Because $P$ and $\Phi$ are diagonal and all constraints act row by row, these assignments can be combined independently.
\end{proof}

Write $H=(H_1,H_0)$, and let $C_i(H)$ be the set of distinct admissible row tuples $\vartheta_i=(W_{i:},M_{i:},\lambda_i,\phi_i)$ constructed above. Thus $|C_i(H)|\leq 2$, where $|\cdot|$ denotes set cardinality. The reduced coefficients describe the evolution of the differences. To determine which candidates generate the original trajectories, we use the affine term omitted by differencing.

\scaleFigure

\subsection{Compatibility and Uniqueness}
\begin{proposition}\label{prop:trajectory}
Let a noiseless trajectory $z(0),\ldots,z(T)$ with integer final time $T\geq 2$ satisfy the reduced affine recurrence~\eqref{eq:t6} with coefficients $(H_1,H_0)$ and the same affine term $g\in\mathbb{R}^n$, given by $g=z(2)-H_1z(1)-H_0z(0)$ at every applicable time. A physically valid candidate decomposition $(P,Q,R,\Lambda,\Phi)$ realizes this trajectory under the original EPO updates for some hidden initial state if and only if
\begin{equation}
g=(I-\Lambda)\{z(1)-Rz(0)\}.
\label{eq:t17}
\end{equation}
For multiple trajectories,~\eqref{eq:t17} must hold for each trajectory with its own affine constant $g^{(k)}$.
\end{proposition}

\begin{proof}
The affine term in~\eqref{eq:t6} is $g=\Phi(I-\Lambda)y(0)$. From the expressed update, $z(1)=\Phi y(0)+Rz(0)$, so $\Phi y(0)=z(1)-Rz(0)$. Because $\Phi$ and $\Lambda$ are diagonal, they commute, and $\Phi(I-\Lambda)y(0)=(I-\Lambda)\Phi y(0)$. Substitution gives~\eqref{eq:t17}. This proves necessity. For sufficiency, define $y(t)=\Phi^{-1}\{z(t+1)-Rz(t)\}$ for $0\leq t\leq T-1$, and let $c=(I-\Lambda)y(0)$. Compatibility gives $g=\Phi c$. Substituting the reduced recurrence and $H_1=P+R$, $H_0=\Phi Q-PR$ then yields $y(t+1)=Py(t)+Qz(t)+c$ for $0\leq t\leq T-2$; diagonality ensures $P\Phi=\Phi P$. The expressed update holds by construction, and the final unobserved private state can be generated by its update. Thus the candidate reproduces every available expressed state.
\end{proof}

After the affine recurrence has been established over the observation interval, each trajectory's compatibility test uses only $z(0),z(1),z(2)$. Later samples provide information for identifying or validating $H$ and the constancy of the affine term. Define the scalar $h_i^{(k)}$ and vector $g^{(k)}\in\mathbb{R}^n$ for trajectory $k$:
\begin{align}
h_i^{(k)}&=z_i^{(k)}(1)-\sum_{j\ne i}(H_1)_{ij}z_j^{(k)}(0),\label{eq:t18}\\
g^{(k)}&=z^{(k)}(2)-H_1z^{(k)}(1)-H_0z^{(k)}(0).\label{eq:t19}
\end{align}
For a row candidate $\vartheta_i$, write $\lambda_i(\vartheta_i)$ for its susceptibility and $r_i(\vartheta_i)=(1-\phi_i)M_{ii}$ for its effective expressed self weight, evaluated at that candidate. Define the surviving candidate set
\begin{equation}
\begin{aligned}
A_i=\bigl\{\vartheta_i\in C_i(H):g_i^{(k)}
  &=(1-\lambda_i(\vartheta_i))\\[-2pt]
  &\quad{}\times[h_i^{(k)}-r_i(\vartheta_i)z_i^{(k)}(0)]\\
  &\qquad\text{for all }k\bigr\}.
\end{aligned}\label{eq:t20}
\end{equation}

\begin{theorem}\label{thm:unique}
Suppose that the noiseless expressed trajectories are generated by a common EPO model satisfying the parameter bounds and row normalization conditions stated in Sec.~\ref{sec:problem} and that these trajectories uniquely determine $H$. All admissible parameter tuples $(W,M,\Lambda,\Phi)$ reproducing these trajectories, with unknown private initial states allowed separately for each trajectory, form the Cartesian product $\prod_{i=1}^n A_i$. The parameter tuple is unique if and only if $|A_i|=1$ for every node.

Fix a generating model with $\lambda_i<1$ for every node. There is a set $\mathscr{E}\subset\mathbb{R}^{2n}$, a union of at most $n$ proper hyperplanes depending only on this model, such that any trajectory with initial pair $(y(0),z(0))\notin\mathscr{E}$ excludes every candidate in $C_i(H)$ different from the generating row, for every node. Consequently, an observation record that uniquely determines $H$ and contains one such trajectory uniquely determines all model parameters. If this trajectory's initial pair has a joint density on $\mathbb{R}^{2n}$, it lies outside $\mathscr{E}$ with probability one.
\end{theorem}

\begin{proof}
Theorem~\ref{thm:reduced} lists all admissible decompositions of $H$, and Proposition~\ref{prop:trajectory} tests exactly which ones reproduce the observations. Since each constraint in~\eqref{eq:t20} involves only its own row candidate, the complete solution set is $\prod_i A_i$. The generating model belongs to this product, so it is nonempty, and uniqueness is equivalent to a singleton in every row.

For the second claim, consider a node with a distinct admissible alternative. Its roots satisfy $p_i\ne r_i$, and the alternative has $p'_i=r_i$, $r'_i=p_i$. Let $a_i=1-\lambda_i>0$ and $a'_i=1-\lambda'_i\geq 0$, where primes denote the alternative. For one trajectory, the generating model gives $h_i=\phi_i y_i(0)+r_i z_i(0)$ and $g_i=a_i\phi_i y_i(0)$; the trajectory superscript is omitted here. The alternative passes compatibility if and only if
\begin{equation}
\phi_i(a'_i-a_i)y_i(0)+a'_i(r_i-p_i)z_i(0)=0.
\label{eq:t21}
\end{equation}
If $a'_i=0$, the coefficient of $y_i(0)$ is $-\phi_i a_i\ne 0$. If $a'_i>0$, the coefficient of $z_i(0)$ is nonzero. Thus~\eqref{eq:t21} defines a proper hyperplane in $\mathbb{R}^{2n}$. The generating model fixes $H$, all candidates, and these coefficients independently of the initial state. Each node has at most one distinct alternative, so $\mathscr{E}$ is the union of at most $n$ such hyperplanes. Outside this union every alternative row is rejected. Conversely, if a single trajectory starts on a hyperplane contributing to $\mathscr{E}$, its corresponding alternative row passes compatibility. Combining this row with the generating rows at all other nodes gives a distinct parameter tuple reproducing the entire trajectory by Proposition~\ref{prop:trajectory}. Thus $\mathscr{E}$ is exactly the set of initial pairs preserving ambiguity for one trajectory among decompositions of the fixed $H$. A finite union of proper hyperplanes has Lebesgue measure zero, giving the probability statement.
\end{proof}

For each row, a distinct alternative survives precisely on its hyperplane~\eqref{eq:t21}. Once $H$ is identified, a single initial pair outside their union eliminates all competing tuples (at most $2^n-1$). Figure~7 in Appendix~A gives two networks whose private edge has opposite signs and whose trajectories coincide on this boundary.

The private network can be determined more fully than the entire parameter tuple. The weight matrix $W$ is unique exactly when all candidates in each $A_i$ have the same private row $W_{i:}$. Its support and sign pattern are unique exactly when those features agree across the surviving candidates, even if their weights or node parameters differ. For any valid candidate,
\begin{equation}
\operatorname{sign}(W_{ij})=\operatorname{sign}\bigl((H_0)_{ij}+p_i(H_1)_{ij}\bigr),\qquad i\ne j.
\label{eq:t22}
\end{equation}
Different root assignments can change private signs while preserving the expressed signs.

When $H$ is uniquely identified and every generating $\lambda_i<1$, a linear calculation provides a sufficient recovery condition. Let $D_i\in\mathbb{R}^{K\times 2}$ have rows $[h_i^{(k)},-z_i^{(k)}(0)]$, and define $a_i=1-\lambda_i\in(0,1)$ and $b_i=a_ir_i\in\mathbb{R}$. Compatibility yields
\begin{equation}
D_i\begin{bmatrix}a_i\\ b_i\end{bmatrix}
=\begin{bmatrix}g_i^{(1)}\\ \vdots\\ g_i^{(K)}\end{bmatrix},\quad
\lambda_i=1-a_i,\quad r_i=b_i/a_i.
\label{eq:t23}
\end{equation}
If $\operatorname{rank}(D_i)=2$ for every node, then $(a_i,b_i)$ is unique and $a_i>0$ permits recovery of $r_i$. The value $p_i=(H_1)_{ii}-r_i$ and Eqs.~\eqref{eq:t12}--\eqref{eq:t16} recover the remaining parameters. Every admissible alternative must solve the same linear equations, which proves uniqueness under this sufficient condition. Since $h_i^{(k)}=\phi_i y_i^{(k)}(0)+r_i z_i^{(k)}(0)$ and $\phi_i>0$, the rank of $D_i$ equals that of the two columns of private and expressed initial values at node $i$ across trajectories. Once $H$ is identified, Eq.~\eqref{eq:t20} resolves the row ambiguity with one generic trajectory when all $\lambda_i<1$.

Based on the above analysis, the proposed procedure for jointly reconstructing the public and private influence networks is summarized as follows:

(1) Fit Eq.~\eqref{eq:t8} by ordinary least squares when $\operatorname{rank}(X)=2n$. One trajectory with $T$ updates contributes $T-2$ rows, so $T\geq 2n+2$ is necessary for this full-rank fit from a single trajectory.

(2) For each trajectory $k$, calculate $h_i^{(k)}$ and $g^{(k)}$ using Eqs.~\eqref{eq:t18} and~\eqref{eq:t19}, respectively. Verify that the affine recurrence in Eq.~\eqref{eq:t6} holds throughout the observed trajectory with the same $g^{(k)}$.

(3) For each node $i$, solve Eq.~\eqref{eq:t11} and assign its roots to $p_i$ and $r_i$. For each distinct assignment, calculate $\lambda_i$, $\phi_i$, and the corresponding rows of $W$ and $M$ using Eqs.~\eqref{eq:t12}--\eqref{eq:t16}. Retain the assignments satisfying the model constraints to form the candidate set $C_i(H)$.

(4) Evaluate the compatibility condition in Eq.~\eqref{eq:t20} for each candidate in $C_i(H)$. A candidate is retained if it satisfies this condition for every observed trajectory. Collect the retained candidates in $A_i$ and repeat this calculation for all nodes.

(5) Compare the corresponding entries of $W$ and $M$ across the candidates in each $A_i$ to determine the uniquely identified edges, signs, and weights. If every $A_i$ contains a single candidate, assemble the recovered rows into $W$ and $M$ and obtain the corresponding individual parameters $\lambda_i$ and $\phi_i$.

The procedure separates reconstruction into the identification of the shared reduced coefficients and the selection of compatible parameter assignments. Each node has at most two candidates, and the compatibility tests are performed independently across nodes. The Cartesian product $\prod_{i=1}^n A_i$ represents all parameter tuples consistent with the observations, while agreement among the retained candidates determines the network properties uniquely supported by the expressed trajectories.

\section{Simulation}

We simulate noiseless expressed opinions on five synthetic graph families and on two empirical signed topologies, Highland Tribes and Bitcoin OTC. Independent private layers test the role of edge overlap. Comparisons with constrained nonlinear least squares use identical expressed trajectories. We also examine sensitivity to mild observation noise using a minimum-residual variant of the compatibility test.

\subsection{Experimental Setup}

Our synthetic experiments span five directed signed network families: Erd\H{o}s--R\'enyi (ER)\cite{r43}, Watts--Strogatz (WS)\cite{r44}, Barab\'asi--Albert (BA)\cite{r45}, and two four-community stochastic block models\cite{r46}, SBM-1 and SBM-2. Under the default partially shared edge mode, the off-diagonal private edge set $E_W$ is generated first---ER draws edges uniformly, WS rewires a directed regular ring with probability 0.25, BA builds high-out-degree hubs via preferential attachment, and SBM-1 and SBM-2 combine dense intra-community and sparse inter-community blocks. A prescribed fraction of $E_W$ is then retained and the remainder replaced by uniformly sampled non-edges to form $E_M$ at a target edge overlap, defined as the ratio of shared edges to the union of both layers, ensuring at least one off-diagonal incoming edge per node in each layer. ER and BA have 35\% negative edges in each layer. In WS, negative edges comprise 15\% of local and 55\% of rewired edges in $W$, and 35\% of all edges in $M$. In SBM-1, edges within each of four communities are positive and edges between communities are negative in both layers. In SBM-2, 10\% of within-community edges are negative in each layer, compared with 85\% of between-community edges in $W$ and 25\% in $M$.

Alongside the partially shared mode, we construct an independent-private control mode in which an expressed network $M$ is generated first from each synthetic or empirical topology and paired with an independently drawn signed ER private network $W$ having 35\% negative edges in synthetic tests, together with a strictly disjoint zero-overlap control enforcing $E_W\cap E_M=\varnothing$. Across all constructions, raw diagonal weights are sampled uniformly from $[0.35,1.0]$, off-diagonal magnitudes from $[0.2,1.0]$, and each row is normalized by its absolute sum as stated in Sec.~II; unless noted otherwise, $\lambda_i$ and $\phi_i$ are drawn uniformly from $[0.2,0.8]$ and initial opinion vectors from $[-1,1]^n$.

We denote our finite-candidate decomposition and trajectory-screening method by EPO-Cand, and benchmark it against row-wise constrained nonlinear least squares (CNLS) on identical expressed trajectories. For each node $i$, CNLS minimizes the squared expressed-opinion prediction residuals over all trajectories and steps by recursively propagating the private state. The unnormalized diagonal weight is fixed to 1, bounded off-diagonal ratios are normalized by absolute row sum to form $W_{i:}$ and $M_{i:}$, parameters $\lambda_i$ and $\phi_i$ are mapped into $(0,1)$ via $0.01+0.98/(1+e^{-u})$, and initial private states are jointly optimized. Off-diagonal ratios lie in $[-30,30]$ and start uniformly in $[-2,2]$, while logistic coordinates $u\in[-6,6]$ and initial private states in $[-3,3]$ start uniformly in $[-1,1]$. Each row is solved from multiple random starts using SciPy's trust-region reflective solver\cite{r47,r48} with at most 500 function evaluations and tolerances of $10^{-11}$ on the objective, parameters, and gradient, keeping the start with lowest fit residual.

Weight recovery accuracy is measured by the relative Frobenius norm error,
\begin{equation}
\operatorname{RelErr}_F(A)=\frac{\lVert\widehat A-A\rVert_F}{\lVert A\rVert_F},\qquad
A\in\{W,M\}.
\end{equation}
For off-diagonal edges in noiseless tests, we apply a fixed threshold $\tau=10^{-8}$ to define the ground-truth edge set $E$ and the recovered edge set $\widehat E$, and quantify edge identification accuracy by the F1 score,
\begin{equation}
\mathrm{F1}=\frac{2\lvert E\cap\widehat E\rvert}{\lvert E\rvert+\lvert\widehat E\rvert}.
\end{equation}
The sign accuracy SignAcc is defined as the proportion of ground-truth edges that are both detected and assigned the correct sign, with missed edges counted as errors. All metrics are computed separately for the private layer $W$ and the expressed layer $M$. Every simulation configuration in Secs.~IV~B and IV~C is evaluated over 30 independent trials, for which we report medians together with individual-trial distributions or interquartile ranges.

\begin{table*}[t]
\centering
\caption{Two-layer reconstruction accuracy on empirical signed topologies and independent-private controls, reported as medians over 30 independent trials per configuration. Bold values mark better performance on Highland Tribes. Edge counts denote off-diagonal directed edges, and EPO-Cand attains $\mathrm{SignAcc}=1.000$ in both layers throughout.}
\label{tab:empirical}
\small
\setlength{\tabcolsep}{3pt}
\begin{tabular}{@{}lrrrrc@{}}
\hline\hline
Network / Construction & Nodes $n$ & Private edges & $\operatorname{RelErr}_F(W)$ & $\operatorname{RelErr}_F(M)$ & Edge F1 ($W$ / $M$) \\
\hline
Highland Tribes, EPO-Cand & 16 & 116 & $\mathbf{4.09\times10^{-14}}$ & $\mathbf{4.22\times10^{-14}}$ & \textbf{1.000 / 1.000} \\
Highland Tribes, CNLS (4 starts) & 16 & 116 & 0.355 & 0.153 & 0.818 / 0.868 \\
\hline
Bitcoin OTC, Top-10 & 10 & 53 & $1.09\times10^{-13}$ & $1.03\times10^{-13}$ & 1.000 / 1.000 \\
Bitcoin OTC, Top-50 & 50 & 979 & $5.63\times10^{-13}$ & $5.13\times10^{-13}$ & 1.000 / 1.000 \\
Bitcoin OTC, Top-200 & 198 & 5,308 & $1.02\times10^{-12}$ & $9.20\times10^{-13}$ & 1.000 / 1.000 \\
Bitcoin OTC, Top-1000 & 988 & 18,202 & $5.22\times10^{-13}$ & $5.94\times10^{-13}$ & 1.000 / 1.000 \\
Bitcoin OTC, Full core & 4,784 & 33,681 & $9.09\times10^{-13}$ & $1.19\times10^{-12}$ & 1.000 / 1.000 \\
\hline
Highland Tribes, public $M$ + random $W$ & 16 & 94 & $4.96\times10^{-14}$ & $4.19\times10^{-14}$ & 1.000 / 1.000 \\
Bitcoin OTC Top-200, public $M$ + random $W$ & 198 & 1,578 & $9.28\times10^{-14}$ & $8.38\times10^{-14}$ & 1.000 / 1.000 \\
\hline\hline
\end{tabular}
\end{table*}

\subsection{Reconstruction Results on Synthetic Networks}

We first evaluate joint two-layer recovery at $n=50$ with mean degree parameter 6 across the five synthetic network families at approximately 70\% inter-layer edge overlap and across the independent-private controls. Each setting uses 30 independent trials with $K=35$ trajectories of $T=6$ updates. EPO-Cand identifies a unique parameter assignment at every node in all 300 trials, with edge F1 and sign accuracy of 1.000 in both layers and relative weight errors near $10^{-13}$ (Fig.~2). These results include the disjoint-support case. Equations (7) and (15) explain why the algebra allows different edge sets: the off-diagonal entries of $H_1$ give the expressed interactions up to row scaling, while $H_0+PR$ gives the scaled off-diagonal private interactions. The measured errors vary across graph families. Here the 35 short trajectories supply 140 regression rows for 100 columns of $X$, jointly estimating $H$ before screening the candidate rows. For comparison, Fig.~7 gives a single-trajectory example that also identifies $H$, and shows how the initial condition determines whether the sign of a private edge is identifiable.

We next compare EPO-Cand with multi-start CNLS on ER networks with mean degree parameter 6, 35\% negative edges, and approximately 60\% edge overlap. At each $n\in\{6,8,10,12,14,16,20,25,30\}$, we run 30 independent trials with $T=5$ and $K=\max\{12,\lceil 2.5n/3\rceil\}$. CNLS uses three random starts at every size, with up to 500 evaluations per start and row. Each row has $2(n-1)$ free weight ratios, two response parameters, and $K$ private initial values, giving $2n+K$ fitted coordinates. In Figs.~3(a) and 3(b), median CNLS weight error jumps from order $10^{-13}$ at $n=8$ to $7.1\times 10^{-2}$ at $n=10$ and reaches 0.66 at $n=30$, while median private-edge F1 falls below 0.4 at $n=30$. These results describe the stated solver and budget. EPO-Cand uses linear regression and row-wise compatibility tests to recover a unique solution in all 270 trials, with edge F1 and sign accuracy of 1.000.

For the scale test, sparse ER and BA networks have mean degree parameter 8, 35\% negative edges, and partially shared supports. Eleven sizes from 10 to 2,000 use $T=5$ and $K=\max\{12,\lceil 2.5n/3\rceil\}$, with 30 independent trials per family at every size, totaling 660 trials. EPO-Cand resolves 658 of 660 networks uniquely. In these 658 trials, edge F1 and sign accuracy are 1.000 in both layers and median weight errors are on the order of $10^{-14}$ to $10^{-12}$ (Fig.~3(c)). The two remaining trials, one BA network at $n=50$ and one ER network at $n=1{,}000$, each have one unresolved row whose two self-weights differ by order $10^{-7}$--$10^{-6}$, giving a discriminant of order $10^{-13}$--$10^{-12}$ in Eq.~(11). These cases illustrate the sensitivity of the root test near coincidence at the compatibility tolerance $10^{-10}$.

\noiseFigure

To test sensitivity beyond the noiseless setting, we perturb the expressed opinions with additive Gaussian observation noise $\widetilde z^{(k)}(t)=z^{(k)}(t)+\eta\sigma_z\xi^{(k)}(t)$, where $\sigma_z$ is the standard deviation of the clean expressed trajectories and $\xi^{(k)}(t)$ has independent standard normal entries. We evaluate seven relative noise levels $\eta\in\{0.05\%,0.1\%,0.2\%,0.5\%,1\%,2\%,5\%\}$ on 50-node ER networks with mean degree parameter 6, 35\% negative edges, and approximately 70\% edge overlap, running 30 independent trials at each noise level. Keeping the ordinary least-squares fit of Eq.~(8) and the algebraic row formulas in Eqs.~(11)--(16) unchanged, we omit the exact equality checks, average the affine term $g^{(k)}$ over $t=2,\ldots,T$, clamp any small negative discriminant in Eq.~(11) to zero, and replace the hard compatibility threshold in Eq.~(20) with minimum-residual selection over the admissible root assignments using the root-mean-square residual of Eq.~(17) with a mild boundary penalty when the estimated susceptibility exceeds 1. We use $K=400$ short trajectories of $T=3$ updates to lower regression variance without late-time signal decay, and raise the support detection threshold for $\widehat E$ to $\tau=0.025$ to filter regression background noise while retaining true nonzero weights.

Figure~4 summarizes the resulting reconstruction accuracy. In Fig.~4(a), median relative weight errors in both layers increase smoothly with $\eta$, from $2.86\times10^{-2}$ for $W$ and $2.75\times10^{-2}$ for $M$ at $\eta=0.05\%$ to 0.22 at $\eta=0.5\%$ and 0.34--0.35 at $\eta=1\%$. In Fig.~4(b), signed structure recovery remains highly accurate throughout the mild-noise regime: for $\eta\le 0.5\%$, both layers retain median edge F1 between 0.988 and 1.000 and median sign accuracy between 0.990 and 1.000. At $\eta=1\%$, the private layer still achieves median $\mathrm{F1}=0.961$ and $\mathrm{SignAcc}=0.969$, compared with 0.924 and 0.986 for the expressed layer. At $\eta=2\%$ and $5\%$, edge F1 drops more rapidly as background noise creates false-positive entries above $\tau$, whereas true-edge sign accuracy degrades more gradually, remaining at 0.875 for $W$ and 0.800 for $M$ at $\eta=5\%$. Because Eq.~(8) regresses on noisy first differences and Eqs.~(12) and (16) sum absolute perturbations across all zero off-diagonal entries, stronger noise introduces both errors-in-variables bias and row-normalization drift.

\topologyFigure
\bitcoinFigure

\subsection{Reconstruction Results on Real-World Network Topologies}

We further evaluate EPO-Cand on two empirical signed social network topologies, with complete results summarized in Table~I. The first dataset is the anthropological Highland Tribes network\cite{r49,r50} retrieved from the KONECT repository\cite{r51}, which records 58 undirected alliance and enmity ties among 16 tribes in the Central Highlands of New Guinea, corresponding to 116 off-diagonal directed edges. In the primary test, the private layer $W$ retains the empirical edge locations and signs, whereas the expressed layer $M$ preserves all alliance ties while dropping or flipping a subset of enmity ties and adding cross-faction links to capture public diplomacy diverging from private stance; individual parameters are drawn from $[0.25,0.78]$ with $K=24$ trajectories of $T=8$ updates. Over 30 independent trials on identical expressed trajectories, EPO-Cand achieves zero-error edge and sign identification with $O(10^{-14})$ median weight error in both layers, whereas four-start CNLS incurs large weight errors and frequent edge misidentifications in both layers.

The second dataset is the directed weighted signed Bitcoin OTC trust network\cite{r52,r53}. Using the latest trust rating between each ordered user pair to form the adjacency matrix $A$, we construct the private layer $W=\operatorname{Norm}(A+I)$ and the expressed layer $M=\operatorname{Norm}(A^\top+I)$, where $\operatorname{Norm}$ denotes row-wise absolute-sum normalization, yielding a strongly asymmetric two-layer signed network whose layers have distinct, mutually transposed edge locations. Five induced subgraphs are extracted by descending total node degree, iteratively retaining nodes with at least one off-diagonal incoming edge in $W$. Each of the five scales is evaluated over 30 independent trials with $T=4$ updates and $K=\max\{16,\lceil 5n/4\rceil\}$ trajectories, totaling 150 trials. The full active core contains 4,784 nodes and 33,681 off-diagonal edges per layer, of which 8.8\% are negative; Fig.~5 shows the Top-50 active core. We also evaluate independent-private controls over 30 independent trials each, pairing Highland Tribes and Bitcoin OTC Top-200 as $M$ with independent signed ER layers $W$ of mean degree 6 with 50\% negative edges and mean degree 8 with 35\% negative edges, respectively. Table~I summarizes these medians; both layers achieve edge F1 and sign accuracy of 1.000 at the precision reported.

On the 4,784-node core, $K=5{,}980$ noiseless trajectories supply five expressed-state observations apiece ($T=4$), so the difference regression has 11,960 rows and 9,568 columns. Figure~6 compares the reconstructed layers with the two matrices constructed from the trust graph. In Fig.~6(a), the median maximum entrywise error grows monotonically from $10^{-13}$ to $10^{-11}$ as edge counts increase from 53 to 33,681, whereas $\operatorname{RelErr}_F(W)$ peaks near $10^{-12}$ on the dense Top-200 hub core and dips slightly at larger scales as sparser peripheral nodes raise self-weights and $\lVert W\rVert_F$ under $L_1$ normalization. Across all 30 full-core trials, all 33,681 off-diagonal private edges are recovered with the correct sign, and the median full-core relative errors are $9.09\times10^{-13}$ for $W$ and $1.19\times10^{-12}$ for $M$ (Table~I), with entrywise errors in Fig.~6(b) concentrated between $10^{-15}$ and $10^{-12}$ and both transposed patterns cleanly separated in Figs.~6(c)--6(f). These figures measure recovery from simulated EPO trajectories on the observed trust topology.

\section{Conclusion}

We characterized which private influence relations can be inferred from noiseless expressed opinions in a signed two-layer model. Once the reduced coefficients are identified, each node admits at most two assignments of its effective self-weights. Full-trajectory compatibility tests these assignments independently. If each individual retains nonzero anchoring to their initial private opinion, a single trajectory selects a unique parameter tuple for almost every joint initial state. In noiseless simulations, the sparse scaling experiment yielded a unique reconstruction in 658 of 660 trials. Under mild observation noise on synthetic networks, replacing exact compatibility screening with minimum-residual selection preserved accurate signed support recovery up to roughly 1\% relative noise. Tests on constructed Highland Tribes and Bitcoin OTC layers also recovered their signed edge patterns, including the 4,784-node Bitcoin OTC core.

The algebraic separation of the two layers also explains why their signed supports can be recovered when their edges differ. The reduced recurrence determines the off-diagonal signs of the expressed layer directly. For the private layer, each surviving assignment determines a row of weights, and agreement across assignments identifies individual edges and signs. The analytic procedure estimates the reduced recurrence by linear regression and checks at most two candidates per row. It yielded smaller reconstruction errors than constrained nonlinear least squares at the larger sizes in the comparison under the reported restart and evaluation budgets.

Several promising directions merit future work. First, combining sparse or errors-in-variables regularized regression with noise-robust compatibility scoring will further extend this framework to stronger noise, finite-length, and partially observed opinion records, such as online discussions and multi-stage collective forecasting experiments\cite{r54}. Second, adapting two-layer identifiability analysis to time-varying, multi-topic, and higher-order social networks, as well as nonlinear opinion updates, will capture richer sociological settings. Finally, incorporating partial structural priors or targeted interventions offers a practical way to lower observation requirements on sparse real-world data.

\section*{Author Declarations}
\subsection*{Conflict of Interest}
The authors declare that they have no known competing financial interests or personal relationships that could have appeared to influence the work reported in this paper.

\section*{Data and Code Availability}
The data that support the findings of this study are available from the corresponding authors upon reasonable request.

\appendix
\section{A two-node identifiability boundary}
\label{app:boundary}
Consider the sign of the private edge $2\to 1$ in a two-node network.
Write each row in the order $(W_{i:},M_{i:},\lambda_i,\phi_i)$.
For node 1 take the generating row and its alternative to be
\[
\begin{aligned}
\vartheta_1 &= \bigl((2/5,-3/5),(1/2,1/2),1/4,1/5\bigr),\\
\vartheta'_1 &= \bigl((20/29,9/29),(1/5,4/5),29/50,1/2\bigr).
\end{aligned}
\]
Both networks use $\vartheta_2=\bigl((1/3,2/3),(-1/4,3/4),2/3,3/5\bigr)$.
The effective self-weights $(p_1,r_1)$ are $(1/10,2/5)$ in the generating
network and $(2/5,1/10)$ in the alternative.
Substitution in Eq.~(7) gives the same $H_1,H_0$ for both networks, although the private weight $W_{12}$ changes from $-3/5$ to $9/29$.

Given the first two public observations $z(0),z(1)$, Eq.~(2) fixes the alternative private initial opinion at $y'_1(0)=[2y_1(0)+3z_1(0)]/5$, with $y'_2(0)=y_2(0)$.
The alternative then matches the first update. Its discrepancy at the second update is
\begin{equation}
z'_1(2)-z_1(2)=\frac{63z_1(0)-33y_1(0)}{500}.
\tag{A1}\label{eq:boundary}
\end{equation}
Thus $21z_1(0)=11y_1(0)$ is exactly the ambiguity line in Eq.~(21) for this row. On the line, the affine terms also coincide, so the two networks generate the same public opinions at every later time. Off the line, the second update distinguishes the signs.

\boundaryFigure

Figure~7 fixes $y_1(0)=0.21$, $y_2(0)=0.8$, and $z_2(0)=-0.5$.
At $z_1(0)=0.11$, the alternative starts from $y'_1(0)=0.15$ and reproduces both public opinions indefinitely. Changing $z_1(0)$ by $\pm 0.2$ gives discrepancies of $\pm 0.0252$ at the second update.
For each of the three trajectories, six updates yield a $4\times 4$ design matrix $X$ of rank four in exact arithmetic. The ambiguity on the line therefore persists even when this single record uniquely determines the reduced coefficients.

\makeatletter
\let\auto@bib@innerbib\@empty
\makeatother

\end{document}